\newcommand{\gev}{\, {\rm GeV}}
\newcommand{\tev}{\, {\rm TeV}}

\newcommand{\xfb}{\, {\rm fb}}
\newcommand{\xpb}{\, {\rm pb}}
\newcommand{\beq}{\begin{equation}}
\newcommand{\eeq}{\end{equation}}
\newcommand{\bea}{\begin{eqnarray}}
\newcommand{\eea}{\end{eqnarray}}

\newcommand{\gsim}{\lower.7ex\hbox{$\;\stackrel{\textstyle>}{\sim}\;$}}
\newcommand{\lsim}{\lower.7ex\hbox{$\;\stackrel{\textstyle<}{\sim}\;$}}

\documentclass[aps,prev,twocolumn,preprintnumbers,%
             floatfix,nofootinbib]{revtex4}
\usepackage{graphicx}
\usepackage{mathrsfs}
\usepackage{amssymb}
\usepackage{verbatim}
\usepackage{color}

\def\stacksymbols #1#2#3#4{\def\theguybelow{#2}
   \def\vp{\lower#3pt}
   \def\sp{\baselineskip0pt\lineskip#4pt}
   \mathrel{\mathpalette\intermediary#1}}

\def\intermediary#1#2{\vp\vbox{\sp
    \everycr={}\tabskip0pt
    \halign{$\mathsurround0pt#1\hfil##\hfil$\crcr#2\crcr
             \theguybelow\crcr}}}

\def\comment#1{}
\def\to{\rightarrow}

\def\xfb{\, {\rm fb}}
\def\u1x{U(1)_X}
\newcommand{\nc}{\newcommand}
\nc{\LL}{L} \nc{\vv}{\tilde{v}} \nc{\ccdot}{\!\cdot\!}
\nc{\gsm}{G_{SM}}
\nc{\vfive}{\mathbf{5}\oplus\mathbf{\overline{5}}}
\nc{\vten}{\mathbf{10}\oplus\mathbf{\overline{10}}}
\nc{\zhol}{Z^{\rm hol}}

\begin{document}

\preprint{~~~~CERN-PH-TH/2010-203; MCTP-10-41; IFT-10-11; DESY 10-143}


\title{Supersymmetric mass spectra for gravitino dark matter \\ with a high reheating temperature}

\author{L.~Covi$^a$, M.~Olechowski$^b$, 
S.~Pokorski$^b$~\footnote{Hans Fischer Senior Fellow, Institute for Advanced 
Studies, Technical University, Munich, Germany}, \\ 
K.~Turzy\'nski$^b$, J.~D.~Wells$^{c,d}$}
\affiliation{
\vspace{0.1cm}
${}^{a}$ Deutsches Elektronen-Synchrotron, DESY, Hamburg, Germany \\
\vspace*{0.1cm}
${}^{b}$Institute of Theoretical Physics, University of Warsaw, Ho\.za 69, 00-681, Warsaw, Poland\\
\vspace*{0.1cm}
${}^{c}$CERN Theory Group (PH-TH), CH-1211 Geneva 23, Switzerland\\
\vspace*{0.1cm}
${}^d$MCTP, University of Michigan, Ann Arbor, MI 48109, USA}

\begin{abstract}

Supersymmetric theories with gravitino dark matter generally do not allow the high reheating temperature required by thermal leptogenesis without running afoul of relic abundance or big bang nucleosynthesis constraints. We report on a successful search for parameter space that does satisfy these requirements. The main implication is the near degeneracy of the gluino with the other neutralinos in the spectrum. The leading discovery channel at the LHC for this scenario is through monojet plus missing energy events.

\end{abstract}

\maketitle

\setcounter{equation}{0}


\section{Gravitinos from High reheating temperature}

Our goal is to investigate the supersymmetry spectrum that allows gravitinos to be the lightest supersymmetric partner (LSP) dark matter of the universe and also allows thermal leptogenesis to explain the baryon asymmetry, all while retaining the successful predictions of big bang nucleosynthesis. We explain each of these in turn, highlighting the potential sources of conflict between them, and finally settling upon an explanation requiring a neutralino next-to-lightest supersymmetric partner (NLSP) degenerate with the gluino, and then investigating its consequences for the Large Hadron Collider (LHC).

To begin we note that in gauge mediated supersymmetry~\cite{GMrev} the 
lightest supersymmetric particle (LSP) is generically the gravitino, 
with mass ranging from the eV scale to tens of GeV. There are many 
nice features of this model that we do not detail here, but suffice it 
to say that it is a powerful and viable organizing principle for the 
superpartner spectrum, and motivates our interest in the gravitino as 
the LSP. The identity of the next to lightest supersymmetric particle 
and the details of the spectrum are model dependent 
(see {\em e.g.},~\cite{Lalak:2008bc}). 
Being the LSP and stable, the gravitino is the leading candidate for dark matter in these models, apart from the possibility of very long-lived particles in the messenger sector \cite{Hamaguchi:2009db}. However, a light gravitino is at most Warm Dark Matter and not favoured by structure formation; indeed, present observations already constrain its mass to be above a few keV if it decouples as a relativistic thermal relic~\cite{Viel:2005qj}.

This is not the only source of gravitinos in the early Universe,
as they do not have to reach thermal equilibrium densities to be cosmologically important.
On one hand, scatterings in the thermal plasma produce gravitinos with the abundance proportional to the reheating temperature after inflation~\cite{Bolz:2000fu,Pradler:2006qh,Rychkov:2007uq}:
\begin{equation} 
\label{c1a}
\Omega_{\tilde G}^\mathrm{TP}h^2 \approx \frac{\left(\frac{T_\mathrm{R}}{10^{9}\,\mathrm{GeV}}\right)\left(\frac{m_\mathrm{NLSP}}{300\,\mathrm{GeV}}\right)^2}{\left(\frac{m_{3/2}}{1\,\mathrm{GeV}}\right)}
\sum_r \gamma_r\cdot\left(\frac{M_r}{m_\mathrm{NLSP}}\right)^2 \, ,
\end{equation}
where $M_r$ denote physical gaugino masses and the coefficients $\gamma_r$ depend on the ratios of the gauge couplings at the reheating scale and  the scale of the physical gaugino masses: with the 1-loop RGE for the gaugino masses, the values of $\gamma_r$ can be 
evaluated for $T_\mathrm{R}=10^9\,(10^7)\,\mathrm{GeV}$ as $\gamma_3=0.48-0.56\,(0.62-0.74)$, $\gamma_2=0.57\,(0.54)$, $\gamma_1=0.22\,(0.17)$, where the range for $\gamma_3$ corresponds to the gluino masses ranging from  200 to 900 GeV \cite{Olechowski:2009bd}. We have only included the 
production of the goldstino component of the gravitino, which dominates for $m_\mathrm{NLSP}/m_{3/2} > \mathcal{O}(10)$. 
On the other hand, gravitinos are also produced in the gravitational decays of the NLSP, but for
$\Omega_\mathrm{NLSP}h^2\ll 1$ or $m_{3/2}/m_\mathrm{NLSP}\ll 1$ these decays are a negligible 
source of gravitino dark matter; moreover, too high a fraction of such a nonthermal and warmish dark 
matter component can cause too much erasure of the cosmic structures at small scales \cite{Jedamzik:2005sx}.
Other contributions to the gravitino abundance can arise from inflaton decay~\cite{Kawasaki:2006gs} or from the reheating process~\cite{Allahverdi:2004si}, but they are more model dependent and we will not discuss them further.

Thus, the gravitino abundance is largely determined directly by the 
reheating temperature $\Omega_{\tilde G}\sim T_R$, as suggested by 
eq.~(\ref{c1a}).  
If this were the only way the reheating temperature affected 
the scenario, one could contemplate a simple explanation for the cold 
dark matter by tuning $T_R$ to achieve the required $\Omega_{\tilde G}$.

However, there are other implications to the choice of $T_R$ that must be considered. Thermal leptogenesis requires a rather high $T_R$ to be successful, which may yield too much dark matter unless the gravitino mass is lifted to higher values ($\Omega_{\tilde G}\sim T_R/m_{3/2}$, assuming $m_\mathrm{NLSP}$ is fixed). However, higher gravitino mass means a slower NLSP decay, which follows a normal thermal relic history and then dumps its decay energy into $\mathrm{NLSP}\to \tilde G+X$ after big bang nucleosynthesis (BBN). BBN compatibility depends on both the number density of the NLSP $n_\mathrm{NLSP}$ and its decay lifetime $\tau_\mathrm{NLSP}$ and other quantities (such as the decay branching fractions, the gravitino mass, etc.). Thus, there are strains and potential incompatibilities when requiring compatibility among gravitino dark matter, thermal leptogenesis and big bang nucleosynthesis.

In the next two subsections we shall describe in more detail the constraints that arise from requiring both successful thermal leptogenesis and compatibility with BBN. We shall then explain in sec.~\ref{sec:NLSP} our emphasis on scenarios with neutralino NLSP. In sec.~\ref{sec:max} we  determine the maximum allowed reheating temperature, which is wanted by thermal leptogenesis, consistent with all the constraints. We show how this value depends on other parameters of the theory. In sec.~\ref{sec:LHC} we discuss the implications of the resulting parameter space for the Tevatron and Large Hadron Collider.   We summarize our conclusions in sec.~\ref{sec:conclude} and make some additional final remarks.


\subsection{Thermal Leptogenesis Requirements}

Generation of the baryon asymmetry through thermal leptogenesis remains a theoretically attractive and experimentally viable possibility, as it only uses particles (right-handed neutrinos) and interactions (neutrino Yukawa couplings) already  present in the seesaw models explaining the smallness of the neutrino masses (for a review, see \cite{DavidsonRev}).
Putting it simply, the lepton asymmetries, subsequently converted into baryon asymmetry, are produced in the CP violating  decays of the lightest right-handed neutrinos; it is usually assumed that these neutrinos had been previously produced in scatterings in the thermal plasma. 
So successful thermal leptogenesis 
gives a lower bound on the mass of the lightest neutrino, $M_{N_1}^\mathrm{(min)}$, in usual seesaw 
models (with hierarchical masses of the right-handed neutrinos), which can be translated into a lower 
bound on the reheating temperature after inflation: $T_R^\mathrm{(min)}\approx M_{N_1}^\mathrm{(min)}/5$ ($T_R^\mathrm{(min)}\approx M_{N_1}^\mathrm{(min)}$) in the so-called strong (weak) washout regime. 

Interestingly, the lower bound $T_R^\mathrm{(min)}=1.9\cdot 10^9\,\mathrm{GeV}$ \cite{Antusch:2006gy} found in the model with the largest possible CP asymmetries and optimal (small) washout approximately coincides with values of $T_R$ obtained in generic models predicting the correct baryon asymmetry, found with the Monte Carlo Markov chain techniques \cite{Davidson:2008pf}. 
However, this bound on the reheating temperature is not an absolute one.  
If the initial conditions for leptogenesis include a thermal distribution 
of the lightest right-handed neutrino, the minimal reheating temperature 
is $2.5\cdot 10^8\,\mathrm{GeV}$ \cite{Giudice:2003jh}. This is the 
uncomfortably high reheating temperature with respect to gravitino abundance 
that we referred to above in the introduction. Such a high $T_R$ puts strain 
on a light gravitino LSP being the dark matter while being consistent with BBN,
 {even if one tries to make it as large as possible by requiring that 
the masses of the gauginos and the NLSP are not too far 
apart~\cite{Olechowski:2009bd}}. 
That consistency is the subject of the next subsection.

Before closing this discussion, we remark on a few caveats to what was said 
above. It has been argued that neglecting quantum effects in the dynamics of 
leptogenesis introduces theoretical uncertainties as large as an order of 
magnitude \cite{Anisimov:2010aq}. Furthermore, the bounds discussed above do 
not apply in models with degenerate masses of the right handed neutrinos, 
see {\em e.g.}, \cite{Flanz:1994yx,Turzynski:2004xy}, or in models with 
large cancellations in the seesaw mass formula~\cite{Hambye:2003rt,Raidal:2004vt}. However, these solutions are of somewhat less interest to us here since 
they involve a degree and type of fine tuning that we wish to try to avoid. 
We also note that the reheating temperature may be lowered in models of 
nonthermal leptogenesis, see {\em e.g.}, 
{\cite{Murayama:1993em,Giudice:2008gu}}, or in soft leptogenesis, 
see {\cite{Grossman:2003jv,D'Ambrosio:2003wy,Hamaguchi:2010cw}}. 
Such models, however, require arranging for additional interactions, 
{\em e.g.} a coupling between the inflaton and a right-handed neutrino 
or a coupling of the Higgs doublet 
to the leptonic component of the messenger field. But in this approach the attractive feature of independence of initial conditions is lost -- another consequence that we wish to avoid here.


\subsection{BBN Consistency}

Since the abundance of dark matter (in our case consisting of gravitinos) is  
$\Omega_{\tilde G} h^2=0.110\pm0.006$ \cite{DM},
substituting the minimal reheating temperature $T_R^\mathrm{(min)}$ discussed above into (\ref{c1a}) shows that
the gravitino mass consistent with the dark matter abundance is at least $\mathcal{O}(1\,\mathrm{GeV})$
in the most optimistic case of the gaugino masses degenerate with $m_\mathrm{NLSP}$.
With such gravitino masses the NLSP lifetime,
\beq
\label{tau}
\tau_\mathrm{NLSP} = \left(5.9\cdot 10^4\, {\rm sec}\right) \left(\frac{m_{3/2}}{1\,\mathrm{GeV}}\right)^2 \left( \frac{100\,\mathrm{GeV}}{m_\mathrm{NLSP}}\right)^5
\eeq
for $m_{3/2}/m_\mathrm{NLSP}\ll1$, 
easily exceeds the duration of the big bang nucleosynthesis.
NLSP decays taking place during or after BBN, can alter its successful predictions if the 
relic abundance of the NLSP and/or the hadronic branching fraction in the NLSP decay is 
large enough \cite{Ellis:1984er,Balestra:1986gg,Kawasaki:1994af,Ellis:1995mr,Ellis:2003dn,
Kawasaki:2004qu,fst1,fst2,Roszkowski:2004jd,Cerdeno:2005eu,Jedamzik:2006,Jedamzik:2007qk}. 
Furthermore, if the NLSP is charged, it can bind with nuclei, which facilitates the production of
${}^6\mathrm{Li}$ \cite{Dimopoulos:1989hk,Pospelov:2006sc,Pradler:2007is}. 
This introduces a tension between successful thermal leptogenesis, which requires $T_R>T_R^{(\mathrm{min})}$, and gravitino Dark Matter.  For a given MSSM spectrum and known 
$\Omega_\mathrm{NLSP} h^2$  this tension can be translated with (\ref{c1a}) and (\ref{tau}) into a lower 
bound on $\tau_\mathrm{NLSP}$ and an upper bound on $m_{3/2}$. 

The BBN bounds are weaker and more easily satisfied for shorter lifetimes, see also \cite{Pradler:2006hh,Choi:2007rh,Steffen:2008bt,Cerdeno:2009ns}, and disappear below 0.1 s. 
Therefore, there has been an effort to identify the NLSP candidates for which the BBN bounds are 
weaker than usual, hence allowing for relatively heavy gravitino DM and a high reheating temperature. Several solutions have been proposed, involving either a reduction of the NLSP relic 
abundance compared to the generic case, see {\em e.g.}, \cite{Pradler:2008qc,Covi:2009bk,Hasenkamp:2010if}, or 
the suppressing of the energy released in the NLSP decay kinematically, thanks to extremely small 
mass splitting between the NLSP and the gravitino LSP \cite{Boubekeur:2010nt}.

The type of NLSP also plays an important role and changes the BBN bounds.
One  of the most studied is the stau: 
it naturally is the NLSP in minimal models of gauge mediation with a large messenger number 
and a high scale of supersymmetry breaking (the latter feature also predicts a heavy gravitino), 
and in its decay few energetic hadrons dangerous to BBN are produced. 
Nevertheless, the stau is charged and it is constrained by bound-state 
effects, so for a  typical stau relic abundance, a reheating temperature 
larger than {$\mathcal{O}(10^8\,\mathrm{GeV})$ is excluded, even 
with a compressed
spectrum of stau and gaugino masses \cite{Olechowski:2009bd}}.
Stau relic density can also be suppressed thanks to a large left-right 
mixing in the stau sector. This effect has been studied in the context of the 
CMSSM and possible reheating temperatures as large as. 
$\sim 10^9\,\mathrm{GeV}$ for $\mu<0$  \cite{Pradler:2008qc} 
and  $\mathcal{O}(10^8\,\mathrm{GeV})$ for $\mu>0$ \cite{Bailly:2009pe}
were found. There is also
a possibility that the stau annihilation cross section is enhanced by a Higgs pole~\cite{Ratz:2008qh}. 
All these solutions require a fine tuning among the soft supersymmetry breaking mass parameters.

Sneutrinos as the NLSP easily evade the BBN bounds even for a high reheating temperature 
suitable for thermal leptogenesis, as long as their masses do not exceed 200-300 GeV
\cite{Kanzaki:2007pd,Kawasaki:2008qe}. 
Hovewer, arranging for a sneutrino NLSP requires a strong degeneracy between the soft supersymmetry
breaking mass parameters for the superpartners of the left- and right-handed leptons 
(see \cite{Olechowski:2009bd} for gauge mediation) or non-universal Higgs masses 
(see  \cite{Buchmuller:2006nx,Covi:2007xj,Ellis:2008as} for gaugino and gravity mediation).

Therefore, our interest turns to a relatively unexplored option, that of thermal leptogenesis with a neutralino NLSP {and a general spectrum of supersymmetric particles},
{as most recent studies on neutralino NLSP were done in the context of CMSSM with a gravitino LSP}. In the next section we shall describe the challenges of this option, then do the quantitative work in sec.~\ref{sec:max} to show that it can work under some circumstances, and then describe the LHC implications in sec.~\ref{sec:LHC}.


\subsection{Leptogenesis with Neutralino NLSP\label{sec:NLSP}}

At first sight, neutralinos as candidates for the NLSP appear much worse than sleptons or
sneutrinos \cite{fst2}.
First, hadrons are often found 
in the decay channels of the neutralinos (roughly, the hadronic branching fraction $B_h$ ranges from 3 to 50 percent for neutralino masses between 100 and 1000 GeV \cite{Covi:2009bk}). Secondly, the bino NLSP usually has a large relic abundance and it has been considered mostly in the context 
of models with universal gaugino masses, {i.e.\ with $M_r/g_r^2$ independent of
the gauge group index $r$ at the scale of supersymmetry breaking}\footnote{At 1 loop, $M_r/g_r^2$ is a renormalization group invariant; for concretness, we impose the universality condition at the boundary scale $10^{15}\,\mathrm{GeV}$.}. With the gluino mass 
approximately 5 times larger than the bino mass at the electroweak scale, it is clear from eq.\ (\ref{c1a}) 
that the resulting reheating temperature is smaller by an order of magnitude with respect to the case 
of little or no hierarchy between the gaugino masses. As will become evident in the figures of sec.~\ref{sec:max}, the  typical maximum reheating
temperature for the bino NLSP with universal gaugino masses reaches only $10^5-10^6\,\mathrm{GeV}$, which is much too low for successful thermal leptogenesis needs.

Reconciling models of gauge mediated supersymmetry breaking with thermal leptogenesis
requires either lowering the leptogenesis temperature or relaxing the BBN bounds.
As discussed above, this can be achieved simply 
by reducing the relic density of the NLSP 
{\em e.g.} by means of coannihilations with more strongly interacting particles (which also evades strong cosmological bounds on the presence of metastable charged  or strongly interacting particles). 
Reducing the NLSP relic density by coannihilations remains largely unexplored phenomenologically, 
expecially in this context, but has been explored for the case of bino-stau coannihilation in the CMSSM~\cite{Ellis:2003dn, Roszkowski:2004jd, Cerdeno:2005eu, Pradler:2006hh,Choi:2007rh,Steffen:2008bt, Bailly:2009pe} or in more general models~\cite{Covi:2009bk}.

In the following, we would like to focus on the most promising case of neutralino/gluino mass degeneracy. 
We determine if the reheating temperatures allowed in such a scenario  are consistent with thermal leptogenesis and what fine tunings may be necessary. This part of our discussion depends only on the 
assumption that gravitinos are the dark matter particles, and we assume arbitrary masses 
of the supersymmetric particles, without imposing constraints from specific theoretical models.
Note that for bino and wino NLSP,  the gaugino masses play a domininant role  in the gravitino and NLSP abundances, and a degeneracy helps to reduce both.
Therefore the degenerate spectrum can be considered the best case scenario. 


\section{Maximizing the reheating temperature\label{sec:max}}

Whether a given set of the parameters leading to gravitino LSP and a neutral NLSP is consistent with the BBN bounds depends mainly on 4 quantities: the NLSP mass, its relic abundance, its hadronic branching fraction and its lifetime (which can be traded for $m_{3/2}$ with the use of (\ref{tau})). Refs.\ \cite{Jedamzik:2006,Jedamzik:2007qk} find the excluded regions on the $(\tau_\mathrm{NLSP},\Omega_\mathrm{NLSP}h^2)$ plane for different values of the hadronic branching fraction and two values of the NLSP masses, 100 and 1000 GeV. Up to a moderate shift in the allowed $\Omega_\mathrm{NLSP}h^2$,  the excluded regions are very similar for both these masses. In order to apply the BBN bounds for a general set of parameters, we interpolate (linearly on a logarithmic scale) between the results of \cite{Jedamzik:2006,Jedamzik:2007qk}, constructing the maximal allowed NLSP abundance,
$\Omega_\mathrm{NLSP}^\mathrm{max}h^2(m_\mathrm{NLSP},\tau_\mathrm{NLSP},B_h)$. Using this function, for a given MSSM spectrum, we can calculate $\Omega_\mathrm{NLSP}h^2$, find the maximal $m_{3/2}$ for which $\Omega_\mathrm{NLSP}h^2<\Omega_\mathrm{NLSP}^\mathrm{max}h^2$ and, from (\ref{c1a}), find the maximal allowed reheating temperature.

In general, $\Omega_\mathrm{NLSP}^\mathrm{max}h^2$ decreases with growing $B_h$, but it is a non-monotonic function of $\tau_\mathrm{NLSP}$: it consists of 4 convex parts, reflecting the bounds for the abundance of ${}^4\mathrm{He}$, $\mathrm{D}$, ${}^6\mathrm{Li}$ and ${}^3\mathrm{He}$.
Since for increasing NLSP mass the predicted $\Omega_\mathrm{NLSP}h^2$ and $B_h$ increase, at certain values of $m_\mathrm{NLSP}$, we may find the BBN-allowed regions (around local maxima of $\Omega_\mathrm{NLSP}h^2$ as a function of $\tau_\mathrm{NLSP}$) close. This makes the bounds for $m_{3/2}$ and $T_R$ discontinuous functions of $m_\mathrm{NLSP}$.

In the case of no coannihilations, we calculate the NLSP
relic density with the computer code {\tt micrOmegas} \cite{Belanger:2006is,Belanger:2008sj}.
In the case with
neutralino/gluino degeneracy we include the coannihilations (taking into account nonperturbative
contributions) using the prescription described in detail in Appendix A.
For hadronic branching fractions for bino, wino and higgsino NLSP we use the results obtained in \cite{Covi:2009bk}, while we take it equal to 1 in the gluino NLSP case. 


\subsection{Bino NLSP}

\begin{figure}
\begin{center}
\includegraphics*[height=7cm]{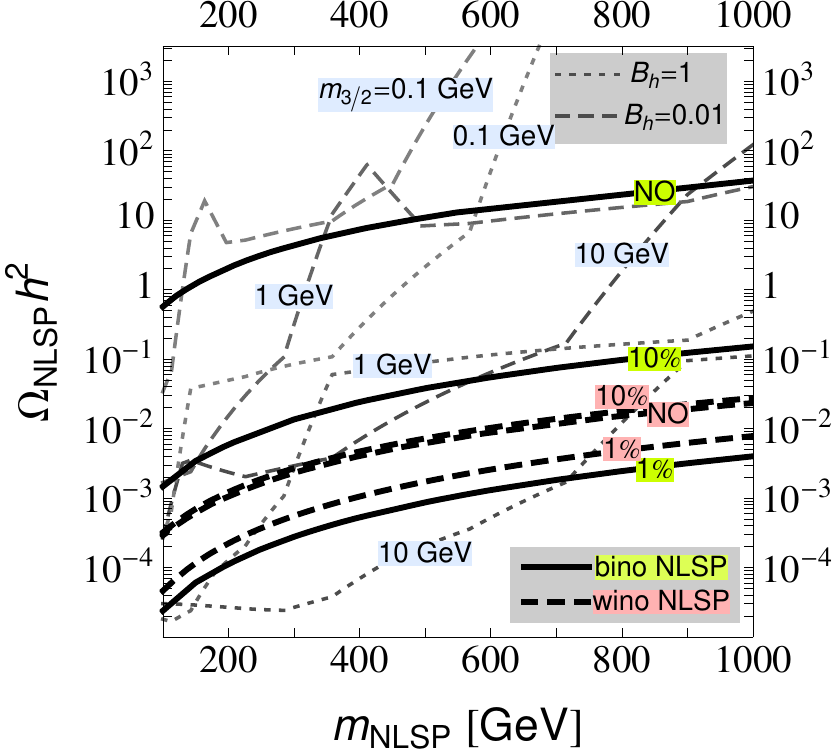}
\end{center}
\caption{Gray dashed lines show
the BBN bounds on $(m_{\tilde B},\Omega_\mathrm{NLSP}h^2)$ plane for three different values
of the gravitino mass, $m_{3/2}=0.1,\,1,\,10\,\mathrm{GeV}$ and two values of $B_h=0.01,\,1$
Also shown are predictions for $\Omega_\mathrm{NLSP}h^2$ 
for the non-degenerate NLSP case (labeled {\em NO}) and with NLSP/gluino degeneracy of 10\% and 1\%; thick black solid (dashed) lines correspond to the bino (wino) NLSP.  \label{fo1}}
\end{figure}

\begin{figure}
\begin{center}
\includegraphics*[height=7cm]{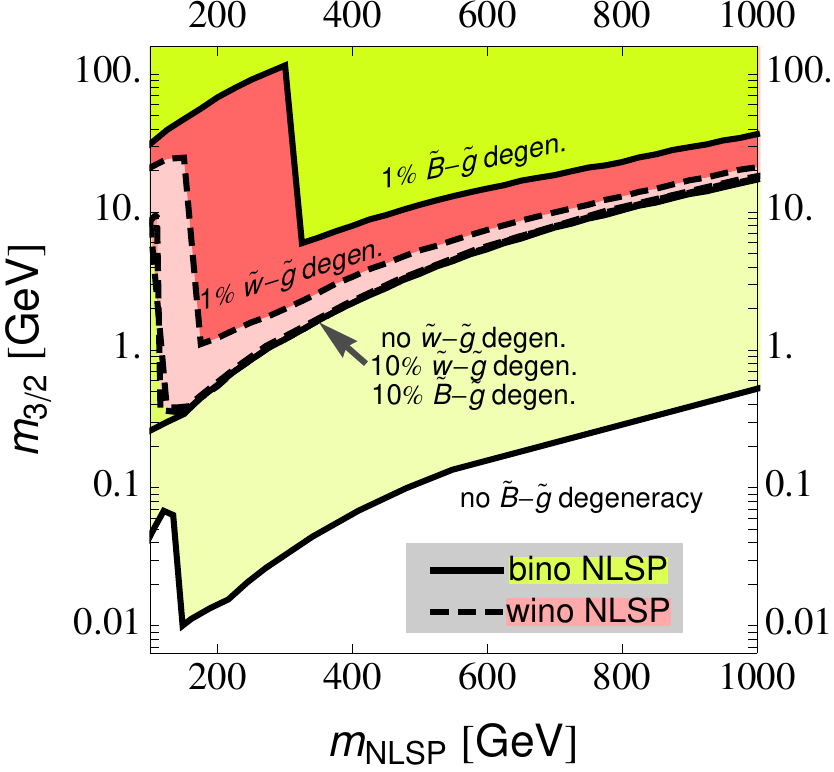}
\end{center}
\caption{BBN bounds on $(m_{\tilde B},m_{3/2})$ plane for the non-degenerate NLSP case (labeled {\em no $\tilde B-\tilde g$ ($\tilde w-\tilde g$) degeneracy}) and with NLSP/gluino degeneracy of 10\% and 1\%; thick black solid (dashed) lines correspond to the bino (wino) NLSP  \label{f1}}
\end{figure}

\begin{figure}
\begin{center}
\includegraphics*[height=7cm]{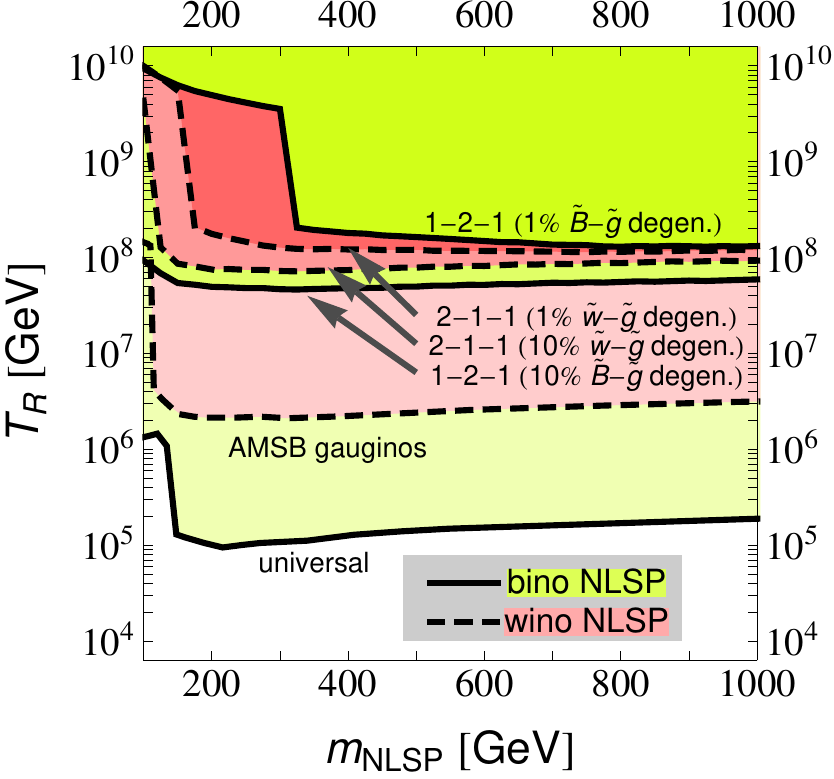}
\end{center}
\caption{BBN bounds on $(m_{\tilde B},T_R)$ plane. Solid lines correspond to the bino NLSP case with universal gaugino masses  and to bino/gluino degeneracy of 10\% and 1\% with $M_1\!:\!M_2\!:\!M_3\approx 1\!:\!2\!:\!1$ (labeled 1-2-1). 
Dashed lines correspond to the wino NLSP case with AMSB gaugino mass spectrum and to wino/gluino degeneracy of 10\% and 1\% with $M_1\!:\!M_2\!:\!M_3\approx 2\!:\!1\!:\!1$ (labeled 2-1-1). 
\label{f2}}
\end{figure}

Our results for bino NLSP are shown on Figures~\ref{fo1}, \ref{f1} and 
\ref{f2}. The maximal allowed $m_{3/2}$, corresponding in practice to 
the largest reheating temperature, can be directly obtained from the BBN
bounds. Figure~\ref{fo1} shows the predictions for 
$\Omega_\mathrm{NLSP}h^2$ for three cases: without degeneracy with gluino, 
with 10\% and 1\% degeneracy, together with the BBN bounds 
for three different values of the gravitino mass, 
$m_{3/2}=0.1,\,1,\,10\,\mathrm{GeV}$ and 
{two values of $B_h=0.01,\,1$}.
In the case of non-degeneracy, it is necessary to know the remaining susy spectrum for the purposes of computing the bino relic abundance. 
{For that, we have chosen the spectrum of minimal gauge mediation
with  one $\mathbf{5}+\mathbf{\bar 5}$ messenger pair and the messenger mass of $10^{15}\,\mathrm{GeV}$. For such a spectrum the main contributions to the bino pair annihilations come
from $t$-channel slepton exchanges and the bino-to-slepton mass ratio is approximately $0.4$.
Since the bino annihilation cross-section is proportional to 
$1/m_{\tilde{\ell}}^4$, by increasing this ratio (e.g. increasing 
the numbers of messengers in minimal gauge mediation),
one can suppress the resulting $\Omega_\mathrm{NLSP}h^2$. However, even with 1\% bino/slepton
mass degeneracy, $\Omega_\mathrm{NLSP}h^2$ is smaller only by a factor of $\sim20$ compared to
our reference case, so it is still larger than what we obtain with 10\% bino/gluino mass
degeneracy.}

Figure~\ref{f1} translates these results into the largest allowed $m_{3/2}$ 
for each of the three cases and Figure~\ref{f2} shows the largest allowed 
reheating temperature for each case. Since the reheating temperature depends 
on the pattern of gaugino masses, we assume universal masses for the case 
without degeneracy and a pattern
$M_1\!:\!M_2\!:\!M_3\approx 1\!:\!2\!:\!1$ for the degenerate cases. 

We see that with 1\% bino-gluino mass degeneracy, we are able to reach a 
reheating temperature as high as a few $10^9\,\mathrm{GeV}$ with 
$m_{\tilde B}<300\,\mathrm{GeV}$.  Even with only $10\%$ bino-gluino mass 
degeneracy, we are able to reach  {$T_R\gsim 0.7\cdot 10^8\gev$} 
for bino masses all the way up to $1\tev$.

\subsection{Wino NLSP}

Our results for wino NLSP are also shown on Figures~\ref{fo1}, \ref{f1} 
and \ref{f2} in the same way as for the bino NLSP case. 
The only difference is instead of the universal gaugino mass pattern,
which always gives the bino as the lightest gaugino, we use the spectrum 
arising in models with anomaly mediated supersymmetry breaking 
(AMSB)~\cite{AMSB1,AMSB2}. This serves as a good illustration of a theory approach 
giving the wino as the lightest gaugino.

Models with nondegenerate wino NLSP allow for a reheating temperature from a few $10^6\,\mathrm{GeV}$ 
for the AMSB gaugino spectrum, with $m_{\tilde g}\sim 10\,m_{\tilde w}$, 
to a few $10^7\,\mathrm{GeV}$ 
for milder mass hierarchies (given by the 10\% $\tilde w-\tilde g$ degeneracy line with almost the same 
$\Omega_\mathrm{NLSP}h^2$ as the nondegenerate case).
Although such $T_R$ is larger than in the models with the generic bino NLSP, 
a degeneracy with a gluino helps less or does not even help at all in 
reducing $\Omega_\mathrm{NLSP}h^2$. 
It can be seen from eq.~(\ref{sigmaeff}) in Appendix A that, in the limit of a very strong degeneracy, 
$(m_{\tilde g}/m_\mathrm{NLSP}-1) \to 0$, and a dominant gluino annihilation cross section, the resulting effective annihilation cross section is proportional to $(16+g)^{-2}$, where $g=2\,(6)$ is the number of the NLSP degrees of freedom for bino (wino) NLSP. Therefore we expect the effective cross-section for
the wino to be smaller than for the bino and the wino abundance to be larger by $ \sim (11/9)^2\approx1.5 $.
Considering as well that the wino hadronic branching ratio is a factor $ \tan^{-2}\theta_W \sim 3.3$ 
larger than for the bino, the stronger bound is explained.
We see that a reheating temperature of a few $10^9\,\mathrm{GeV}$ is also possible in the
wino NLSP case with a 1\% wino-gluino mass degeneracy, but due to the above effects
such large reheating temperatures are possible only in the low mass region, 
for $m_{\tilde w}<200\,\mathrm{GeV}$. Note that a light wino window, with masses around 100 GeV,
is present also for the pure wino case, without coannihilations with the gluinos \cite{Covi:2009bk}.

\begin{figure}
\begin{center}
\includegraphics*[height=7cm]{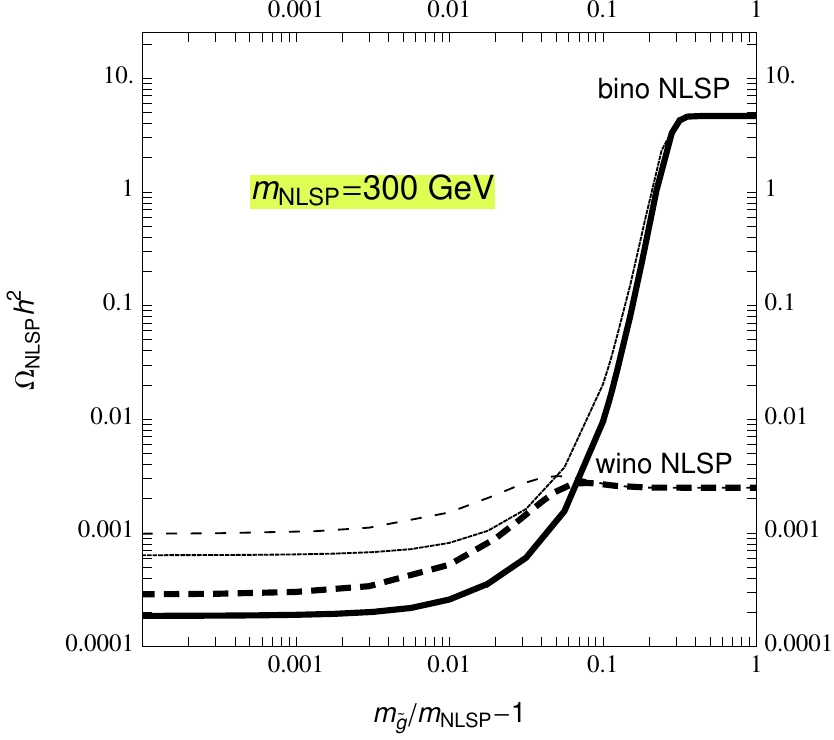}
\end{center}
\caption{NLSP relic density calculated in the presence of gluino coannihilations
as a function of the degeneracy parameter $m_{\tilde g}/m_\mathrm{NLSP}-1$ for $m_\mathrm{NLSP}=300\,\mathrm{GeV}$. Solid (dashed)
lines correspond to bino (wino) NLSP. Thick (thin) lines show the results with (without)
nonperturbative contributions. 
\label{g1}}
\end{figure}

We also note that a 10\% wino/gluino degeneracy actually increases the NLSP relic density despite a larger annihilation cross section for gluinos, as it can be seen in Figure \ref{g1} as a little `bump' in the predictions 
for $\Omega_\mathrm{NLSP}h^2$ in the wino NLSP case.
This happens due to the presence of the `weights' $\gamma_i^2$ in (\ref{sigmaeff}), obeying $\gamma_0+\gamma_{\tilde g}=1$, and the fact that the real coannihilation cross-section involving a wino and a gluino 
in the initial state is negligible for larger squark masses.
In this case, as discussed in the appendix A, the effective cross-section reaches a minimum when 
the ratio of the gluino over wino weights is equal to the ratio of the wino over gluino cross-sections.
The increase in the abundance is at most $ 1 + \sigma_0/\sigma_{\tilde g} $, so it is negligible
for the bino case, but visibile for the wino.


\subsection{Higgsino NLSP}

The case of higgsino NLSP in models of gauge mediation is more involved. 
Although arranging for cancellations between various contributions to the 
soft sypersymmetry breaking Higgs mass parameter $m_{H_2}^2$, and hence 
$\mu$ parameter, is most welcome phenomenologically, it requires some fine 
tuning in the boundary conditions.
Furthermore, for light neutralinos, the hadronic branching fraction is 
quite sensitive to the gaugino admixture and, {\em e.g.}, 5\% bino content 
can lower $B_h$ even by an order of magnitude \cite{Covi:2009bk}.
{Moreover a non-vanishing gaugino fraction opens up also the channel
of resonant annihilation via the pseudoscalar Higgs so that the higgsino
number density can vary strongly around $ 2 m_{\tilde h} \sim m_A $. }
For these reasons, we only indicate here the predictions for the 
higgsino NLSP in the conservative case of no resonant annihilation.
With the gauginos twice heavier than the higgsino, we obtain the maximal reheating temperature {$(2-3)\times10^7\,\mathrm{GeV}$ for higgsino masses between 100 and 1000 GeV and $5\%$ bino admixture}.
Larger reheating temperatures are surely possible if the annihilation 
proceeds on the resonance, at the cost of a fine-tuning between the 
neutralino and Higgs masses \cite{Covi:2009bk}. 

For the case of a higgsino NLSP, coannihilation with the gluinos is perhaps less natural since
one could expect all the gauginos to be much heavier than the higgsinos. Also the equilibrium
between higgsinos and gluino mainly proceeds through small Yukawa couplings and is somewhat
less effective than for gauginos. 
{Finally, one needs to know the full spectrum of the higgsinos to account for the effective number
of the degrees of freedom participating in coannihilations. This can vary from 8, if the mass splittings among the higgsinos are not larger than with the gluino, to 2, if the mass gap between the lightest higgsino and the gluino is much smaller than the mass splittings among the higgsinos. The former case
requires decoupling of heavy binos and winos: a situation with an extreme fine-tuning of the input parameters at the high scale of supersymmetry breaking and, as we have already seen in the wino NLSP case, a large number of coannihilating states leads to larger $\Omega_\mathrm{NLSP}h^2$; hence, the latter case appears more plausible.}

We explored nevertheless the possibility of higgsino/gluino coannihilations 
within models of general gauge mediation with the messenger scale $10^{15}\,\mathrm{GeV}$.
We found an intermediate result between
bino and wino neutralino: the annihilation cross-section lies between those for bino and
wino NLSP, and in the limit of strong NLSP/gluino degeneracy one higgsino participates in coannihilations much more efficiently than the others. {For a $300\,\mathrm{GeV}$ higgsino,
we obtain the relic density of a few $10^{-4}$, while at the Higgs resonance it goes down to a few $10^{-5}$.}


\subsection{Gluino NLSP}

Although gluino NLSP has a small relic density, with $\Omega_\mathrm{NLSP}h^2$ ranging from $6\cdot 10^{-5}$ to $2\cdot 10^{-3}$ for gluino masses from 200 to 1000 GeV, successful primordial nucleosynthesis places very stringent limits on the presence of long-lived strongly interacting relic particles after the BBN \cite{Kusakabe}. 
If the NLSP is coloured, its lifetime should not exceed 300 sec and this is the origin of the constraint 
$T_R<3-7\cdot 10^7\,\mathrm{GeV}$ for $m_{\tilde g}$ ranging from 200 to 1000 GeV and the bino
and the wino twice as heavy as the gluino.
We also note here that the gluino NLSP relic density is smaller than the relic
density of a stop with the same mass \cite{Berger:2008ti}, hence the maximal allowed reheating
temperature in the stop NLSP case is lower than for the gluino NLSP.


\section{LHC Discussion\label{sec:LHC}}

Taking into consideration the requirements of CP asymmetries, BBN, and gravitino dark matter, we have concluded that an interesting approach to viable leptogenesis leads to nearly degenerate gauginos and a spectrum of scalar superpartners somewhat higher in mass. In particular, the gluino mass needs to be rather close in value to the NLSP mass such that coannihilation effects can sufficiently reduce the number density of the NLSP so as to not disrupt BBN when the NLSPs decay.  This generic feature of the spectrum has important consequences for LHC discovery of supersymmetry.  

\subsection{Gluino pair production}

If the gauginos are much lighter than the scalars, then the largest  supersymmetry production cross section at the LHC is $\tilde g\tilde g$. In the limit of a nearly exact degeneracy of the neutralino NLSP and gluino masses, there are no visible particles to trigger on from the production and decay of gauginos, since the 
final state is just an invisible neutralino and a very soft gluon or $ q \bar q $ pair.
For extreme degeneracy below a percent or two, the gluino decay is strongly suppressed and a displaced vertex at $ \geq 1\, {\rm mm} $ distances is possible, but again generally with no visible particle to trigger on, only one or two very soft jets.

\begin{figure}
\begin{center}
\includegraphics*[height=7cm]{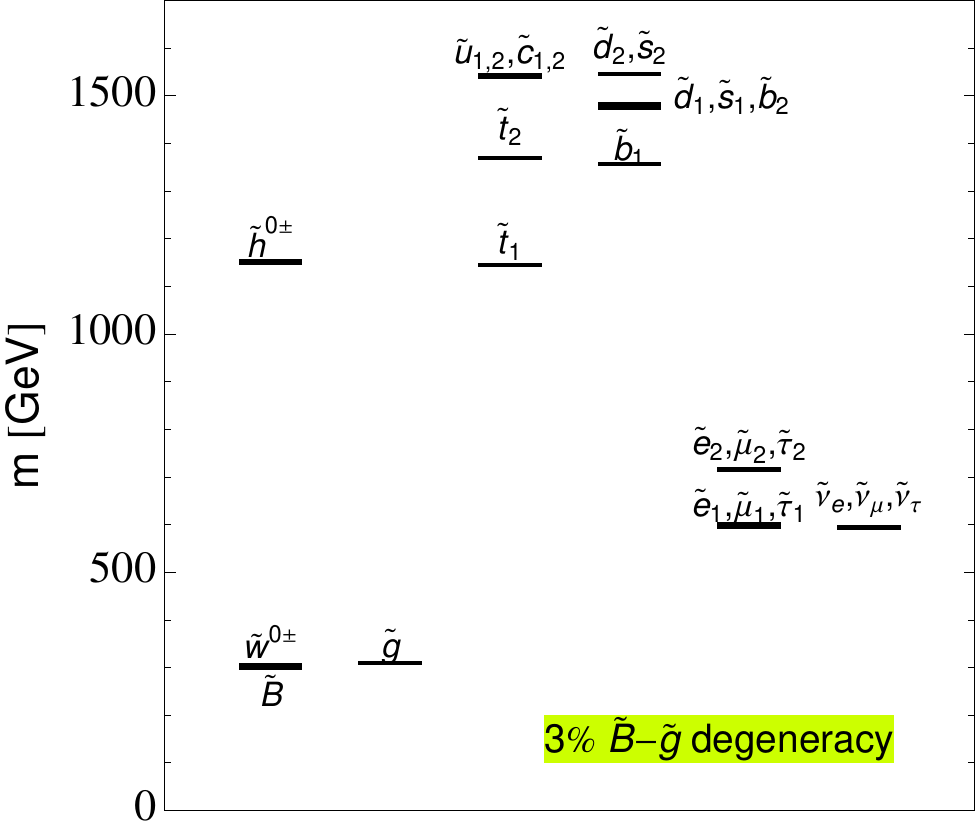}
\end{center}
\caption{Mass spectra of sparticles in the example discussed in the text. The lowest neutralino
line corresponds to $m_{N_1}=299.8\,\mathrm{GeV}$, $m_{N_2}=305.6\,\mathrm{GeV}$ and $m_{C_1}=305.6\,\mathrm{GeV}$, while $m_{\tilde g}=308.9\,\mathrm{GeV}$. The Higgsino mass parameter is $m_{\tilde H} =1153\gev$ and is irrelevant to the ensuing discussion.  \label{fsp1}}
\end{figure}

\begin{table}
\begin{tabular}{|c|c|}
\hline
channel & branching fraction \\
\hline
$\tilde g\to N_1g$ & 0.59 \\
$\tilde g\to N_1q\bar q$ & 0.35 \\
$\tilde g\to N_2g$ & 0.03 \\
$\tilde g\to N_2q\bar q$ & 0.02 \\
\hline
$N_2\to N_1\nu\bar\nu$ & 0.41 \\
$N_2\to N_1\gamma$ & 0.31 \\
$N_2\to N_1\ell^+\ell^-$ & 0.08 \\
\hline
\end{tabular}
\caption{Branching fractions relevant for collider analysis. \label{tabbf}}
\end{table}

For illustration of the issues of detectability of supersymmetric particles at the LHC,
we shall take a closer look at the model with mass spectrum shown in 
Figure~\ref{fsp1}. This model
can be realized within the framework of general gauge mediation with the messenger
scale of $10^{15}\,\mathrm{GeV}$ and $\tan\beta=10$. Bino/gluino degeneracy is 3\% and the
resulting maximal reheating temperature attainable in this model is $3\cdot 10^8\,\mathrm{GeV}$. 
From the branching fractions in table~\ref{tabbf} we see that over $1/3$ of the $\tilde g\tilde g$ events will be in the most advantageous channel $\tilde g\tilde g\to gN_1gN_1$, or in other words, two jets plus missing energy.

For our example model point, the gluino mass is $309\gev$ and the leading order total cross-section at $\sqrt{s}=14\tev$ LHC is $\sigma(\tilde g\tilde g)\simeq 255\pm 5\xpb$.  Thus, in a few $\xfb^{-1}$ of data we expect quite a large number of events from $\tilde g\tilde g$. However, it will be difficult to trigger on these events and discern them above a large background. To see this, we note that in the rest frame of the gluino the energy of the gluon is fixed to be
$ E_g=(m_{\tilde g}^2-m_{N_1}^2)/(2m_{\tilde g})=9.0\gev$. 

\begin{figure}
\begin{center}
\includegraphics*[height=7cm]{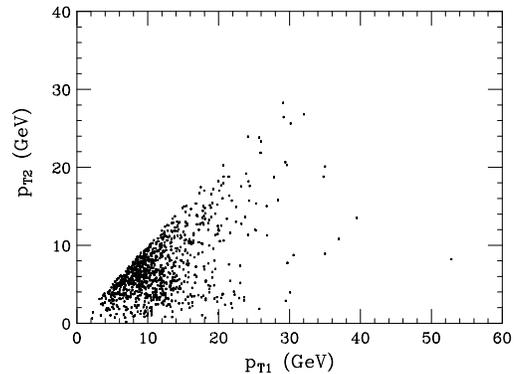}
\end{center}
\caption{Scatter plot of $p_T$ values for the two gluons in $\tilde g\tilde g\to gN_1gN_1$ for 1000 events with $m_{\tilde g}=308\gev$ and $m_{N_1}=300\gev$. This simulation is for a $pp$ collider with $\sqrt{s}=14\tev$ center of mass energy.  \label{fcoll1}}
\end{figure}

We have conducted a MadGraph~\cite{Alwall:2007st} Monte Carlo simulation of the production of gluino pairs followed by decay into gluon plus neutralino. The results are given in Figure~\ref{fcoll1}, where we have plotted the $p_T$ values of each jet at the parton level. Each event is a point in the $(p_{T_1},p_{T_2})$ plane, where $p_{T_1}$  is the higher $p_T$ of the two jets.  The gluon energy in the lab frame may increase or decrease depending on its relative decay direction to the boost direction.  For this reason, the highest $p_T$ jet can be rather large -- in this simulation of 1000 events, the highest $p_T$ obtained was nearly $60\gev$. 

Unfortunately, the visible energy and missing energy is not enough to trigger the events for saving at the LHC.  There is too much background to open the trigger to events of this kinematic topology without significant prescaling that loses the signal. For example, for $\sqrt{s}=7$ TeV collisions,  where the LHC is currently running, there are several triggers that are of potential relevance to gluino gluino production followed by decays to soft things~\cite{Flaecher}; however, each fails to capture a significant number of signal events. There is the ``single jet" trigger requiring $p_T\gsim 110\gev$, the ``dijet trigger" which requires the average $p_T>70\gev$, the ``sum $p_T$" trigger which requires $p_T^{\rm sum}({\rm jets})>200\gev$, and the non-prescaled ``missing $E_T$" trigger requiring ${\rm MET}>60\gev$.  These are only for the $7\tev$ collider, and each of these numbers will be approximately doubled for the $14\tev$ collider. Thus, none of these triggers will efficiently record these events, and  we seek a better path to discovery. 

\begin{figure}
\begin{center}
\includegraphics*[height=7cm]{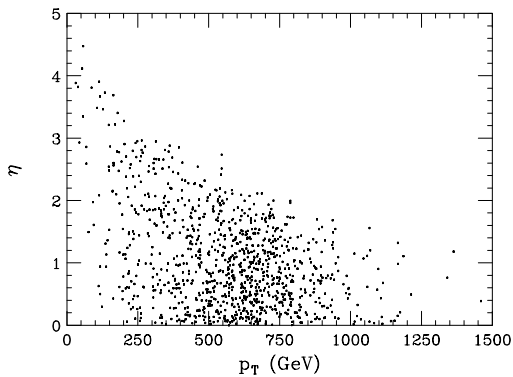}
\end{center}
\caption{Scatter plot of 1000 events at $\sqrt{s}=14\tev$ $pp$ collider from $\tilde q\tilde g\to q\tilde g\tilde g$, where $m_{\tilde g}=300\gev$ and $m_{\tilde q}=1.5\tev$.  The meaning of $\tilde q$ is all first and second generation  squarks: $\tilde u_L$, $\tilde u_R$, $\tilde d_L$, $\tilde d_R$, $\tilde s_L$, $\tilde s_R$, $\tilde c_L$, $\tilde c_R$.  Each event is characterized by the absolute value of the pseudorapidity $\eta$ of the quark jet from $\tilde q\to q\tilde g$ decays, and the $p_T$ of that jet. Given that the gluino is nearly degenerate with all other gauginos including the NLSP, it acts like a source of missing energy. Thus, the $p_T$ of the quark jet is approximately the missing energy of the event also.   \label{fcoll2}}
\end{figure}

When gaugino degeneracy is present, we have seen that the final states in the process are too soft to trigger on. Therefore, we need another process that can generate much higher $p_T$ jets or leptons.   We remark that there is also the prospect of detecting gluino pair production via tagging from an initial state radiation (ISR) jet.  This is a technique that has been advocated for many new physics scenarios that have no substantial visible energy from the final state~\cite{ISR studies}. ISR jets tend to be soft, and backgrounds are determined to large degree by how well the detector is understood and how well fake rates and jet energy measurements can be controlled. We do not pursue this approach here, but merely remark that it could be a useful signal for discovery, or even confirming a model if discovery is made through another channel. It would be especially important in the case of very high degeneracy such that the gluino decay has a displaced vertex, as discussed earlier in the section.


\subsection{Squark-gluino monojet signature}

We wish to determine if there might yet be another signature of value.
Upon inspection of the spectrum it is evident that squarks are too heavy to pair produce efficiently, and certainly the same goes for the more weakly coupled sleptons.  However, the squarks are often not too heavy for the promising signature of a single squark being produced in association with a (much lighter) gluino.

An important contributing factor to the viability of this signature is the 
high parton luminosities for $gq$ initial states that produce $\tilde g\tilde q$ at tree-level. Furthermore, the kinematics of these events are favorable where the large mass difference between the squark and gluino generates a large $p_T$ jet from $\tilde q\to \tilde g q$ decays. There is a also a large amount of missing energy recoiling against this jet. The signature is that of monojet plus missing energy. This has been studied recently within the context of the Tevatron for a wide class of models~\cite{Alwall:2008va}, as well as studies dedicated to the LHC, as we shall discuss below. {Gluino pair production with ISR jet discussed in the previous section can also contribute to this signature, although the jet $p_T$ is typically much softer. We shall ignore that additional small contribution to the high $p_T$ signal defined in the following analysis.}

The model that we simulate is a slightly simplified version of our spectrum that we considered above. We consider here $m_{\tilde g}=300\gev$ and take the light squark masses to be $1.5\tev$ in order to illustrate the signature with a reasonable spectrum. Generically, and certainly in this case,  over 90\% of light squark decays are into $\tilde q\to q\tilde g$, so single jet plus missing energy (i.e., soft-invisible decays of gluinos) becomes the most important signature to consider. The total leading order production cross-section at the 14 TeV pp LHC collider is $3.4\xpb$. Although this is nearly two orders of magnitude below the $\tilde g\tilde g$ cross-section, the picobarn rate is high enough to record many thousands of events in the course of a few inverse femtobarns of integrated luminosity. Thus, it is a promising signature. In Figure~\ref{fcoll2} 1000 events are simulated, where we give the pseudorapidity ($\eta$) and the $p_T$ values for each simulated event.  To reduce backgrounds and increase reliability of the analysis, it is often required that the jet be central ($|\eta|<2$) and have large $p_T$ ($p_T>200\gev$). Those two requirements still leave the vast majority of events available for analysis.

\begin{figure}
\begin{center}
\includegraphics*[height=7cm]{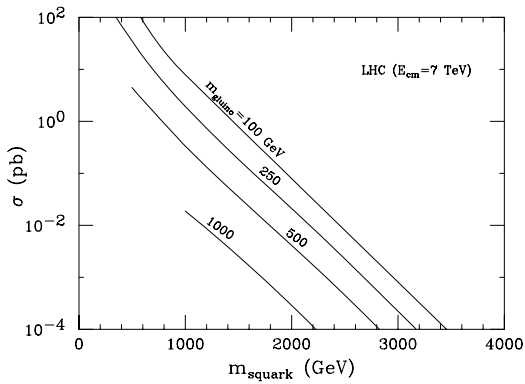}
\end{center}
\caption{Squark gluino production cross-section at $\sqrt{s}=7\tev$ $pp$ collider as a function of squark mass for various values of the gluino mass. The squark masses refer to first two generation squarks.  \label{fcoll3a}}
\end{figure}

\begin{figure}
\begin{center}
 \end{center}
\includegraphics*[height=7cm]{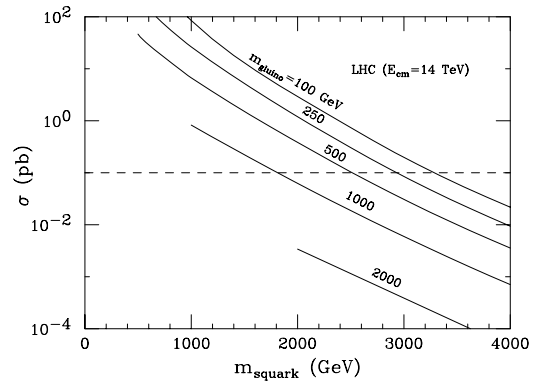}
\caption{Squark gluino production cross-section at $\sqrt{s}=14\tev$ $pp$ collider as a function of squark mass for various values of the gluino mass. The squark masses refer to first two generation squarks. 
The dashed line of $\sigma=0.1\xpb$ is the estimated lower limit of the cross-section that the LHC would be sensitive to when seeking a BSM contribution after more than $1\xfb^{-1}$ of data is accumulated. 
 \label{fcoll3b}}
\end{figure}

Figures~\ref{fcoll3a} and~\ref{fcoll3b} plot the total squark plus gluino production cross-section at  a $7\tev$ and $14\tev$ center of mass energy $pp$ collider as a function of the squark mass (first two generations) for various values of the gluino mass. 
When the squark mass is much greater than the gluino mass, the resulting jet $p_T$ from squark decays to gluino plus quark are very high, and we get a strong single jet plus missing energy signature with very high trigger efficiency.

The question of what the background is for the single jet plus missing energy is notoriously subtle. Our purposes here are to describe the basic features of the background, and give an estimate of expectations. Only a full detector simulation after careful engagement with the LHC data can ultimately determine what precise sensitivity levels can be reached.

Nevertheless, we can compare our signal to the background after cuts advocated in table 2 of~\cite{CMSjet14TeV}. First, the majority of our signal will be one jet plus missing energy. Multiple jets will arise from higher order corrections, which increases the signal; however, we do not include these, thereby losing out on small additional signal, but also not being affected adversely by the ``Number of jets $<3$" cut implemented by~\cite{CMSjet14TeV}. Nevertheless, we know that that cut is very important in reducing $t\bar t$ and QCD backgrounds, and we can assume with overall impunity to the signal that it has been applied to the background.  We can also apply $E_T^{\rm miss}>400\gev$, which, given that we are working to leading order, automatically also implies the simultaneously required $p_T(jet)>350\gev$.  It is automatic because at our leading order computation $p_T(jet)=E^{\rm miss}_T$ to a good approximation.  We can also require that the pseudo-rapidity of our jet is less than 1.7 as~\cite{CMSjet14TeV} requires.  The remaining two azimuthal angle cuts in table 2 of~\cite{CMSjet14TeV} have no consequence to us because they are automatically satisfied in our approximation. Furthermore, they have little effect on the larger backgrounds anyway.

The combination of $E_T^{\rm miss}>400\gev$ and $\eta_{jet}<1.7$ tends to 
reduce our signal by only $\sim 30\%$ which is within the uncertainty of 
QCD corrections and other uncertainties of the analysis. For the background, 
if we combine all these cuts we get  approximately less than 50 events 
per $100\xpb^{-1}$ of data. In other words, the background is about 
$0.5\xpb$.  This estimate is also consistent with the results of 
refs.~\cite{Vacavant:2001sd,Rizzo:2008fp}.  Furthermore, these backgrounds 
can be measured well since at high missing $E_T$ they are dominated by 
$Z(\nu\bar\nu)+j$ which can be normalized to the $Z(l^+l^-)+j$ rate. 
Thus, assuming accurate computations and measurements can be done for 
the background, and uncertainties can dip below $20\%$, we estimate 
reaching sensitivities down to approximately the $0.1\xpb$ level for 
the signal. In $1\xfb^{-1}$ of data, for example, we would have about 
100 signal events at that cross-section, and results would be very near 
threshold for detectability. Thus, we can tentatively conclude that 
the $0.1\xfb$ cross-section line is approximately the threshold for 
discovery of BSM singlet jet plus missing energy signature at $14\tev$ LHC. 
From that determination, everything above the dashed $1\xpb$ line in 
Figure~\ref{fcoll3b} may be detectable at the LHC with over $1\xfb^{-1}$ 
of data. 
Generically this leads to discovery sensitivity of gluinos less than 
$\sim $ TeV with squarks above even a few TeV.

\subsection{Tevatron Remarks}

The parameter space we have been considering is where the gauginos, including gluino, are nearly degenerate. We should remark on what bounds there might be at the Tevatron for this scenario, if any. The limits on the gluino mass from CDF and D0 experiments are usually quoted to be above $\sim 300\gev$~\cite{PDG}; however, those limits assume that the gluino decays into an LSP of much lower mass such that the visible energy in the event is very high and can be triggered on. That is not the case in our model.  The true capability of a limit  should be much lower.

When the gluino mass is degenerate with the LSP to within 10 GeV, the Tevatron has the potential to exclude its existence up to 160 GeV when the lumonisity exceeds $2\xfb^{-1}$~\cite{Chen:1996ap}.  The signal is gluino pair production in association with a jet, leading to a monojet plus missing energy signature. Signal to background requirements are unlikely to allow improvement beyond 160 GeV even with greater luminosity~\cite{Chen:1996ap}. It has been suggested~\cite{Gunion:1999jr} that the photon plus missing energy signature could be more probing than the monojet plus missing energy signature when luminosity exceeds $1\xfb^{-1}$. Although the signal is lower, the signal to background is higher for any given gluino mass, and with sufficient luminosity the higher signal to background signature always wins in sensitivity.  The probing sensitivity for gluinos nearly degenerate with LSP in the photon plus missing energy channel with greater than $1\xfb^{-1}$ has been estimated to be 175 GeV~\cite{Gunion:1999jr}. Further discrimination from background may be possible in some cases if gluino decays frequently produce leptons, albeit very soft ones, through intermediate chargino or neutralino decays to the LSP.  We are not aware of any Tevatron experimental paper that has reported an analysis for these scenarios, so the numbers above are suggestive of what could be done, and not what has been achieved yet.


\section{Summary \& Conclusions\label{sec:conclude}}

In the case of a gravitino LSP and DM, we have explored the parameter space 
of neutralino NLSP with a nearly degenerate gluino and we found that the 
BBN constraints are strongly relaxed in the bino case, so that a reheating 
temperature sufficiently high for thermal leptogenesis  and the right gravitino
abundance are obtained just with a mass degeneracy of the order of a few 
percent between bino and gluino.
So in general a compressed gaugino spectrum helps in reconciling leptogenesis 
with gravitino DM {and neutralino NLSP}.
On the other hand, for the wino and higgsino cases, the coannihilation with 
gluinos does not improve much the situation: for the wino case, it extends 
a bit the light wino window already found in~\cite{Covi:2009bk} for a pure 
wino, while for the higgsino it does not beat the abundance suppression that 
can be {obtained through the resonant annihilation. 
We found also that these degenerate gaugino scenarios can be embedded 
in general gauge mediation with a moderate tuning of the parameters,} 
but since our results depend mostly only on the gaugino masses they are 
not restricted to these specific case and can be extended to any 
supersymmetry breaking scenario {that allows for degenerate 
gaugino masses}.

In all these cases though, the gluinos are {not only nearly degenerate 
with the NLSP, but also quite light $\leq 300 $ GeV} and are therefore 
copiously produced at the LHC. 
{The fact that gravitino DM and thermal leptogenesis together 
give an upper bound on the gluino mass around the TeV scale was 
well-known \cite{Fujii:2003nr}, but in our case in order to suppress 
the NLSP number density we need also a light NLSP and gluino.}
Unfortunately, {since the degeneracy between them is small, 
even if the production cross-section is large}, the main signal of 
a highly energetic track is missing and it is not so simple to 
disentangle the scenario from the QCD background.
We suggested a couple of final states and discuss if they could pass most 
of the LHC trigger cuts and some possible strategies for detection. 
The most promising channel is the gluino squark associated production, 
that gives a highly energetic jet in the final state and could be accessible 
even with  $1\xfb^{-1}$ of data during the early phase of running of the LHC.

\noindent
{\it Acknowledgments:}
We thank A. De Roeck, H. Flaecher, M. Nojiri for useful discussions.

SP thanks the Institute for Advanced Studies at TUM, Munich, for its support 
and hospitality. KT acknowledges the hospitality of the Theory Division at 
CERN and the MCTP.

LC acknowledges the support of the Deutsche Forschungsgemeinschaft under 
the Collaborative Research Centre 676. 
MO and SP acknowledge partial support by the European Research and Training
Network  (RTN) grant `Unification in the LHC era' (PITN-GA-2009-237920) and 
by the  MNiSZW scientific research grant   N N202 103838 (2010 - 2012).
KT is partially supported by the Foundation for Polish Science through 
its programme Homing.


\section*{Appendix A: gluino coannihilations with Sommerfeld enhancement}

Here we discuss how to include nonperturbative effects in the gluino annihilation, often referred to as Sommerfeld enhancement. Our calculation closely follows \cite{Baer,Feldman}.

With the NLSP of mass $m$, the freeze-out temperature is given by:
\beq
x_F = \ln\left[c(c+2)\sqrt{\frac{45}{8}}\frac{g_\mathrm{eff}}{2\pi^3}\frac{1}{\sqrt{g_\ast x_F}} m M_P \langle\sigma v\rangle\right]\, ,
\eeq
where $x=m/T$, $c$ is the coefficient in the relation $Y=(1+c)Y^\mathrm{eq}$ (which we set to 1/2), $M_P$ is the Planck mass (not reduced), $g_\mathrm{eff}$ is the effective number of the degrees of freedom decoupling at freeze-out and $\langle\sigma v\rangle$ is the average containing the annihilation cross-section $\sigma$. The average is:
\beq
\langle \sigma v\rangle = \frac{1}{8m^4TK_2^2\!\left(\frac{m}{T}\right)}\int_{4m^2}^\infty \sigma \,s^{3/2}\beta^2 K_1\!\left(\frac{\sqrt{s}}{T}\right)\,\mathrm{d}s
\eeq
with $\beta=\sqrt{1-4m^2/s}$.

We assume that the annihilation cross-section is the sum of the NLSP annihilation cross-section and the gluino coannihilation cross section:
\beq
\label{sigmaeff}
\sigma = \gamma_0^2 \sigma_0 + \gamma_{\tilde g}^2 \sigma_{\tilde g\tilde g}\, ,
\eeq
where $\gamma_0=g_0/g_\mathrm{eff}$, $\gamma_{\tilde g}=1-\gamma_0$,
\beq
g_\mathrm{eff} = g_0+16 (1+\delta)^{3/2}e^{-\delta x}\, , 
\eeq
$\delta=m_{\tilde g}/m_\mathrm{NLSP}-1$ and $g_0$ is the number of the 
degrees of freedom of the NLSP {and equals 2 (6) for bino (wino) NLSP}. 
This approximation is justified in the limit of heavy squarks, necessary 
to satisfy the Higgs mass bound with light gluinos. In this case in fact 
the coannihilation channel $ \tilde \chi + \tilde g \rightarrow q \bar q $
is strongly suppressed by the large squark masses and can be neglected.
{Note on the other hand that the processes in which a NLSP scatters 
inelastically off a SM particle to become a gluino are efficient down to 
the freeze-out temperature, as the suppression of the cross-section due to 
heavy squarks is compensated by a large abundance of the light SM states.}

The neutralino annihilation cross-section includes many channels. 
For the gaugino neutralino case, an important annihilation channel at low neutralino masses is the 
one into leptons via intermediate sleptons, which are in our case much lighter than the squarks.
This annihilation rate is given by
\bea
\sigma(\tilde \chi \tilde \chi \to \ell \bar{\ell}) &=& 
\frac{3 \alpha |A_{L/R}|^2}{64 \pi \cos^2\theta_W \beta_{\tilde \chi}^2 s} 
\left[ \beta_{\tilde \chi}+\frac{4 \beta_{\tilde \chi} \Delta^2}{(1+ 2 \Delta)^2 - \beta_{\tilde \chi}^2}
\right. \nonumber\\
&+& \vspace*{-0.5cm}
\left. \ln\left(\frac{1+ 2\Delta-\beta_{\tilde \chi}}{1+2 \Delta+\beta_{\tilde \chi}}\right)
\left( 2 \Delta + \frac{1-\beta_{\tilde \chi}^2 }{2(1 + 2 \Delta)} \right) \right] \; ,
\nonumber\\
\eea
where $ \Delta = (m_{\tilde \ell}^2- m_{\tilde\chi}^2)/s $ and
\bea
A_L &=& N_{j1} \pm N_{j2} \cot \theta_W  \quad \mbox{ for } \ell^{-}_L/\nu_{L\ell} \\
A_R &=&  - 2 N_{j1} \; .
\eea
For larger masses and especially for the wino case, also the channel into 
$W^+ W^- $ becomes important and for the {mixed} case the resonant 
annihilation through the Higgs s-channel. In general in the case of wino 
neutralino also the coannihilation with the other charged winos is important 
and the cross-section is not very much smaller than the gluino annihilation
cross-section.
{Since we consider spectra for which the sfermions are generally heavier than the superpartner fermions, we use an approximation 
$\sigma_0\sim \beta/s$ (
$\sigma_0\sim 1/s\beta$) for bino (wino) NLSP, for which the $p$-wave ($s$-wave) annihilation dominates;
the numerical coefficient is chosen so that it 
gives a correct relic density in the absence of coannihilations.}
{In the case of the wino NLSP, one needs, in principle, to consider 
coannihilations between the wino states, but these states are typically 
extremely degenerate in mass, so $\sigma_0$ should be considered an effective 
cross section with a common $\gamma_0$ factor for all wino states 
in (\ref{sigmaeff}). }

The gluino annihilation cross-section reads $\sigma_{\tilde g\tilde g}=\sigma(\tilde g \tilde g\to gg)+\sum_q\sigma(\tilde g \tilde g\to q\bar q)$, where leading contributions, mediated by the gluons, are:
\bea
\label{sgg}
\sigma(\tilde g \tilde g\to gg) &=& \frac{3\pi\alpha_s^2}{16\beta_{\tilde g}^2s} \left[\ln\left(\frac{1+\beta_{\tilde g}}{1-\beta_{\tilde g}}\right)(21-6\beta_{\tilde g}^2-3\beta_{\tilde g}^4) -\right. \nonumber\\
&& \left. \phantom{aaaaaa}-33\beta_{\tilde g}+17\beta_{\tilde g}^3\right] \\
\sigma(\tilde g \tilde g\to q\bar q) &=& \frac{\pi\alpha_s^2\beta_q}{16\beta_{\tilde g}s}(3-\beta_{\tilde g}^2)(3-\beta_q^2)\, ,
\label{sqq}
\eea
with $\beta_a=\sqrt{1-4m_a^2/s}$. 
The contribution of the cross-section with intermediate squarks is negligible in this case.

We see from the expression of the effective cross-section 
eq.~(\ref{sigmaeff}), that the effective numbers of degrees of freedom 
$\gamma $ change {strongly} with the temperature: defining as a variable
the ratio $ z = \gamma_{\tilde g}/\gamma_0 $ it is easy to see that
\beq
\label{sigmaeff2}
\sigma = \sigma_0 \frac{1 + z^2 \sigma_{\tilde g\tilde g}/\sigma_0 }{(1 + z)^2} \, ,
\eeq
and one can show that the fraction on the r.h.s. has minimum as a function of $z$ at 
$ \bar z = \sigma_0/ \sigma_{\tilde g\tilde g} $ and that the minimal value is given as
\beq
\label{sigmaeff-min}
\sigma = \sigma_0 \frac{1}{1 + \bar z} =  
\sigma_0 \frac{  \sigma_{\tilde g\tilde g} }{\sigma_{\tilde g\tilde g} + \sigma_0 } \, .
\eeq
So it is easy to see that this reduction of the cross-section is usually negligible for the
case of bino neutralino, since $ \sigma_0 \ll \sigma_{\tilde g\tilde g} $, but not for the wino.

The Sommerfeld enhancement is accounted for by multiplying each of the cross-sections 
(\ref{sgg}) and (\ref{sqq}) by the
factor
\beq
E_i = \frac{\frac{C_i\pi\alpha_s}{\beta_{\tilde g}}}{1-\mathrm{exp}\left(-\frac{C_i\pi\alpha_s}{\beta_{\tilde g}}\right)}\, ,
\eeq
where $C_i=3/2\,(1/2)$ for the $gg$ ($q\bar q$) final state. The strong coupling constant is evaluated at scales $\beta_{\tilde g}m_{\tilde g}$. This corresponds to averaging over the annihilating states, which produces a smaller correction than summing up all the contributions. We also neglect the possibility of gluinos forming bound states, which would further enhance gluino annihiliations (often quite dramatically). Therefore, our assumption about including the nonperturbative effects in the gluino annihilations 
ensures that the NLSP relic density calculated here is at worst an upper bound, so the claim that
a given model satisfies the BBN bound is robust \cite{Berger:2008ti}.

We can now calculate the final NLSP abundance from:
\beq
Y_\mathrm{NLSP} = \frac{1}{\int_{x_F}^\infty \sqrt{\frac{\pi g_\ast}{45}}\frac{mM_P}{x^2} \langle \sigma v\rangle\,\mathrm{d}x }\, ,
\eeq
which can be used to evaluate the standard cosmological parameter
\beq
\Omega_{\mathrm{NLSP}}h^2 = \frac{ms_0Y_\mathrm{NLSP}}{\rho_c}\, , 
\eeq
where the present entropy density is $s_0=2889.2\,\mathrm{cm}^{-3}$ and the critical density is
$\rho_c=1.0539\cdot 10^{-5}\,\mathrm{GeV\,cm}^{-3}$.


\begin{thebibliography}{99}


\bibitem{GMrev}
For a review, see, e.g.,  G.~F.~Giudice and R.~Rattazzi,
 Phys.\ Rept.\  {\bf 322} (1999) 419
 [arXiv:hep-ph/9801271].

   \bibitem{Lalak:2008bc}
 Z.~Lalak, S.~Pokorski and K.~Turzynski,
 JHEP {\bf 0810} (2008) 016
 [arXiv:0808.0470 [hep-ph]].

 \bibitem{Hamaguchi:2009db}
 K.~Hamaguchi, E.~Nakamura, S.~Shirai and T.~T.~Yanagida,
 JHEP {\bf 1004} (2010) 119
 [arXiv:0912.1683 [hep-ph]].

  \bibitem{Viel:2005qj}
  A.~Boyarsky, J.~Lesgourgues, O.~Ruchayskiy and M.~Viel,
 JCAP {\bf 0905} (2009) 012
 [arXiv:0812.0010 [astro-ph]].

\bibitem{Bolz:2000fu}
M.~Bolz, A.~Brandenburg and W.~Buchmuller,
Nucl.\ Phys.\  B {\bf 606} (2001) 518
[Erratum-ibid.\  B {\bf 790} (2008) 336]
[arXiv:hep-ph/0012052].

\bibitem{Pradler:2006qh}
J.~Pradler and F.~D.~Steffen,
Phys.\ Rev.\  D {\bf 75} (2007) 023509
[arXiv:hep-ph/0608344].

\bibitem{Rychkov:2007uq}
V.~S.~Rychkov and A.~Strumia,
Phys.\ Rev.\  D {\bf 75} (2007) 075011
[arXiv:hep-ph/0701104].

 \bibitem{Olechowski:2009bd}
 M.~Olechowski, S.~Pokorski, K.~Turzynski and J.~D.~Wells,
 JHEP {\bf 0912} (2009) 026
 [arXiv:0908.2502 [hep-ph]].

\bibitem{Jedamzik:2005sx}
K.~Jedamzik, M.~Lemoine and G.~Moultaka,
JCAP {\bf 0607} (2006) 010
[arXiv:astro-ph/0508141].

\bibitem{Kawasaki:2006gs}
 M.~Kawasaki, F.~Takahashi and T.~T.~Yanagida,
 Phys.\ Lett.\  B {\bf 638} (2006) 8
 [arXiv:hep-ph/0603265].

\bibitem{Allahverdi:2004si}
 R.~Allahverdi, A.~Jokinen and A.~Mazumdar,
 Phys.\ Rev.\  D {\bf 71} (2005) 043505
 [arXiv:hep-ph/0410169].

 \bibitem{DavidsonRev}
S.~Davidson, E.~Nardi and Y.~Nir,
Phys.\ Rept.\  {\bf 466} (2008) 105
[arXiv:0802.2962 [hep-ph]].

\bibitem{Antusch:2006gy}
S.~Antusch and A.~M.~Teixeira,
JCAP {\bf 0702} (2007) 024
[arXiv:hep-ph/0611232].

 \bibitem{Davidson:2008pf}
 S.~Davidson, J.~Garayoa, F.~Palorini and N.~Rius,
 JHEP {\bf 0809} (2008) 053
 [arXiv:0806.2832 [hep-ph]].


\bibitem{Giudice:2003jh}
G.~F.~Giudice, A.~Notari, M.~Raidal, A.~Riotto and A.~Strumia,
Nucl.\ Phys.\  B {\bf 685} (2004) 89
[arXiv:hep-ph/0310123].

\bibitem{Anisimov:2010aq}
 A.~Anisimov, W.~Buchmuller, M.~Drewes and S.~Mendizabal,
 Phys.\ Rev.\ Lett.\  {\bf 104} (2010) 121102
 [arXiv:1001.3856 [Unknown]].

\bibitem{Flanz:1994yx}
M.~Flanz, E.~A.~Paschos and U.~Sarkar,
Phys.\ Lett.\  B {\bf 345} (1995) 248
[Erratum-ibid.\  B {\bf 382} (1996) 447]
[arXiv:hep-ph/9411366].

\bibitem{Turzynski:2004xy}
K.~Turzynski,
Phys.\ Lett.\  B {\bf 589} (2004) 135
[arXiv:hep-ph/0401219].

\bibitem{Hambye:2003rt}
T.~Hambye, Y.~Lin, A.~Notari, M.~Papucci and A.~Strumia,
Nucl.\ Phys.\  B {\bf 695} (2004) 169
[arXiv:hep-ph/0312203].

\bibitem{Raidal:2004vt}
M.~Raidal, A.~Strumia and K.~Turzynski,
Phys.\ Lett.\  B {\bf 609} (2005) 351
[Erratum-ibid.\  B {\bf 632} (2006) 752]
[arXiv:hep-ph/0408015].

\bibitem{Murayama:1993em}
  H.~Murayama and T.~Yanagida,
  Phys.\ Lett.\  B {\bf 322} (1994) 349
  [arXiv:hep-ph/9310297].
  
 \bibitem{Giudice:2008gu}
 G.~F.~Giudice, L.~Mether, A.~Riotto and F.~Riva,
 Phys.\ Lett.\  B {\bf 664} (2008) 21
 [arXiv:0804.0166 [hep-ph]].

\bibitem{Grossman:2003jv}
  Y.~Grossman, T.~Kashti, Y.~Nir and E.~Roulet,
  Phys.\ Rev.\ Lett.\  {\bf 91} (2003) 251801
  [arXiv:hep-ph/0307081].
  
\bibitem{D'Ambrosio:2003wy}
  G.~D'Ambrosio, G.~F.~Giudice and M.~Raidal,
  Phys.\ Lett.\  B {\bf 575} (2003) 75
  [arXiv:hep-ph/0308031].
  
\bibitem{Hamaguchi:2010cw}
 K.~Hamaguchi and N.~Yokozaki,
 arXiv:1007.3323 [hep-ph].

\bibitem{DM}
E.~Komatsu {\it et al.}  [WMAP Collaboration],
 Astrophys.\ J.\ Suppl.\  {\bf 180} (2009) 330
 [arXiv:0803.0547 [astro-ph]].

\bibitem{Ellis:1984er}
J.~R.~Ellis, D.~V.~Nanopoulos and S.~Sarkar,
Nucl.\ Phys.\  B {\bf 259} (1985) 175.

\bibitem{Balestra:1986gg}
F.~Balestra {\it et al.},
Nuovo Cim.\  A {\bf 92} (1986) 139.

\bibitem{Kawasaki:1994af}
M.~Kawasaki and T.~Moroi,
Prog.\ Theor.\ Phys.\  {\bf 93} (1995) 879
[arXiv:hep-ph/9403364].

\bibitem{Ellis:1995mr}
J.~R.~Ellis, D.~V.~Nanopoulos, K.~A.~Olive and S.~J.~Rey,
Astropart.\ Phys.\  {\bf 4} (1996) 371
[arXiv:hep-ph/9505438].

\bibitem{Ellis:2003dn}
 J.~R.~Ellis, K.~A.~Olive, Y.~Santoso and V.~C.~Spanos,
 Phys.\ Lett.\  B {\bf 588} (2004) 7
 [arXiv:hep-ph/0312262].

\bibitem{Kawasaki:2004qu}
M.~Kawasaki, K.~Kohri and T.~Moroi,
Phys.\ Rev.\  D {\bf 71}, 083502 (2005)
[arXiv:astro-ph/0408426].

\bibitem{fst1}
J.~L.~Feng, S.~f.~Su and F.~Takayama,
 Phys.\ Rev.\  D {\bf 70} (2004) 063514
 [arXiv:hep-ph/0404198].

\bibitem{fst2}
J.~L.~Feng, S.~Su and F.~Takayama,
 Phys.\ Rev.\  D {\bf 70} (2004) 075019
 [arXiv:hep-ph/0404231].

\bibitem{Roszkowski:2004jd}
 L.~Roszkowski, R.~Ruiz de Austri and K.~Y.~Choi,
 JHEP {\bf 0508} (2005) 080
 [arXiv:hep-ph/0408227].

\bibitem{Cerdeno:2005eu}
 D.~G.~Cerdeno, K.~Y.~Choi, K.~Jedamzik, L.~Roszkowski and R.~Ruiz de Austri,
 JCAP {\bf 0606} (2006) 005
 [arXiv:hep-ph/0509275].

\bibitem{Jedamzik:2006}
K.~Jedamzik,
 Phys.\ Rev.\  D {\bf 74} (2006) 103509
 [arXiv:hep-ph/0604251].

\bibitem{Jedamzik:2007qk}
 K.~Jedamzik,
 JCAP {\bf 0803}, 008 (2008)
 [arXiv:0710.5153 [hep-ph]].


\bibitem{Dimopoulos:1989hk}
S.~Dimopoulos, D.~Eichler, R.~Esmailzadeh and G.~D.~Starkman,
Phys.\ Rev.\  D {\bf 41} (1990) 2388.

 \bibitem{Pospelov:2006sc}
M.~Pospelov,
Phys.\ Rev.\ Lett.\  {\bf 98} (2007) 231301
[arXiv:hep-ph/0605215].

\bibitem{Pradler:2007is}
J.~Pradler and F.~D.~Steffen,
Phys.\ Lett.\  B {\bf 666} (2008) 181
[arXiv:0710.2213 [hep-ph]].


\bibitem{Pradler:2006hh}
J.~Pradler and F.~D.~Steffen,
Phys.\ Lett.\  B {\bf 648} (2007) 224
[arXiv:hep-ph/0612291].

\bibitem{Choi:2007rh}
 K.~Y.~Choi, L.~Roszkowski and R.~Ruiz de Austri,
 JHEP {\bf 0804} (2008) 016
 [arXiv:0710.3349 [hep-ph]].

  \bibitem{Steffen:2008bt}
F.~D.~Steffen,
Phys.\ Lett.\  B {\bf 669} (2008) 74
[arXiv:0806.3266 [hep-ph]].

\bibitem{Cerdeno:2009ns}
D.~G.~Cerdeno, Y.~Mambrini and A.~Romagnoni,
 JHEP {\bf 0911} (2009) 113
 [arXiv:0907.4985 [hep-ph]].

\bibitem{Pradler:2008qc}
J.~Pradler and F.~D.~Steffen,
Nucl.\ Phys.\  B {\bf 809} (2009) 318
[arXiv:0808.2462 [hep-ph]].

\bibitem{Covi:2009bk}
 L.~Covi, J.~Hasenkamp, S.~Pokorski and J.~Roberts,
 JHEP {\bf 0911} (2009) 003
 [arXiv:0908.3399 [hep-ph]].

\bibitem{Hasenkamp:2010if}
 J.~Hasenkamp and J.~Kersten,
 arXiv:1008.1740 [hep-ph].

 \bibitem{Boubekeur:2010nt}
 L.~Boubekeur, K.~Y.~Choi, R.~R.~de Austri and O.~Vives,
 arXiv:1002.0340 [Unknown].

 \bibitem{Bailly:2009pe}
 S.~Bailly, K.~Y.~Choi, K.~Jedamzik and L.~Roszkowski,
 JHEP {\bf 0905} (2009) 103
 [arXiv:0903.3974 [hep-ph]].

\bibitem{Ratz:2008qh}
 M.~Ratz, K.~Schmidt-Hoberg and M.~W.~Winkler,
 JCAP {\bf 0810} (2008) 026
 [arXiv:0808.0829 [hep-ph]].


\bibitem{Kanzaki:2007pd}
T.~Kanzaki, M.~Kawasaki, K.~Kohri and T.~Moroi,
Phys.\ Rev.\  D {\bf 76} (2007) 105017
[arXiv:0705.1200 [hep-ph]].

\bibitem{Kawasaki:2008qe}
M.~Kawasaki, K.~Kohri, T.~Moroi and A.~Yotsuyanagi,
Phys.\ Rev.\  D {\bf 78} (2008) 065011
[arXiv:0804.3745 [hep-ph]].

\bibitem{Buchmuller:2006nx}
 W.~Buchmuller, L.~Covi, J.~Kersten and K.~Schmidt-Hoberg,
 JCAP {\bf 0611} (2006) 007
 [arXiv:hep-ph/0609142].

\bibitem{Covi:2007xj}
L.~Covi and S.~Kraml,
JHEP {\bf 0708} (2007) 015
[arXiv:hep-ph/0703130].

\bibitem{Ellis:2008as}
J.~R.~Ellis, K.~A.~Olive and Y.~Santoso,
JHEP {\bf 0810} (2008) 005
[arXiv:0807.3736 [hep-ph]].

\bibitem{Belanger:2006is}
G.~Belanger, F.~Boudjema, A.~Pukhov and A.~Semenov,
Comput.\ Phys.\ Commun.\  {\bf 176} (2007) 367
[arXiv:hep-ph/0607059].

\bibitem{Belanger:2008sj}
G.~Belanger, F.~Boudjema, A.~Pukhov and A.~Semenov,
Comput.\ Phys.\ Commun.\  {\bf 180} (2009) 747
[arXiv:0803.2360 [hep-ph]].

\bibitem{AMSB1}
 L.~Randall, R.~Sundrum,
 Nucl.\ Phys.\  {\bf B557}, 79-118 (1999).
 [hep-th/9810155].

\bibitem{AMSB2}
 G.~F.~Giudice, M.~A.~Luty, H.~Murayama, R.~Rattazzi,
 JHEP {\bf 9812}, 027 (1998).
 [hep-ph/9810442].


\bibitem{Kusakabe}
M.~Kusakabe, T.~Kajino, T.~Yoshida and G.~J.~Mathews,
 Phys.\ Rev.\  D {\bf 80} (2009) 103501
 [arXiv:0906.3516 [hep-ph]].

\bibitem{Berger:2008ti}
C.~F.~Berger, L.~Covi, S.~Kraml and F.~Palorini,
JCAP {\bf 0810} (2008) 005
[arXiv:0807.0211 [hep-ph]].

\bibitem{Alwall:2007st}
 J.~Alwall {\it et al.},
 JHEP {\bf 0709}, 028 (2007)
 [arXiv:0706.2334 [hep-ph]].


 \bibitem{Flaecher}
 We wish to thank H. Flaecher for discussions on triggers and analysis criteria at the LHC.

 \bibitem{ISR studies}
 For a recent example, see 
 G.~F.~Giudice, T.~Han, K.~Wang and L.~T.~Wang,
 Phys.\ Rev.\  D {\bf 81}, 115011 (2010)
 [arXiv:1004.4902 [hep-ph]].


\bibitem{Alwall:2008va}
 J.~Alwall, M.~P.~Le, M.~Lisanti and J.~G.~Wacker,
 Phys.\ Rev.\  D {\bf 79}, 015005 (2009)
 [arXiv:0809.3264 [hep-ph]].


\bibitem{CMSjet14TeV}
CMS Collaboration, ``Search for Mono-jet Final States from ADD Extra Dimensions," CMS PAS EXO-08-011 (7 August 2009).

\bibitem{Vacavant:2001sd}
 L.~Vacavant and I.~Hinchliffe,
 J.\ Phys.\ G {\bf 27}, 1839 (2001).


\bibitem{Rizzo:2008fp}
 T.~G.~Rizzo,
 Phys.\ Lett.\  B {\bf 665}, 361 (2008)
 [arXiv:0805.0281 [hep-ph]].


\bibitem{PDG}
K. Nakamura et al. (Particle Data Group), ``Review of Particle Physics", J. Phys. G37, 075021 (2010).

\bibitem{Chen:1996ap}
 C.~H.~Chen, M.~Drees, J.~F.~Gunion,
 Phys.\ Rev.\  {\bf D55}, 330-347 (1997).
 [hep-ph/9607421].

\bibitem{Gunion:1999jr}
 J.~F.~Gunion, S.~Mrenna,
 Phys.\ Rev.\  {\bf D62}, 015002 (2000).
 [hep-ph/9906270].

\bibitem{Fujii:2003nr}
  M.~Fujii, M.~Ibe and T.~Yanagida,
  Phys.\ Lett.\  B {\bf 579} (2004) 6
  [arXiv:hep-ph/0310142].

\bibitem{Baer}
 H.~Baer, K.~m.~Cheung and J.~F.~Gunion,
 Phys.\ Rev.\  D {\bf 59} (1999) 075002
 [arXiv:hep-ph/9806361].

\bibitem{Feldman}
D.~Feldman, Z.~Liu and P.~Nath,
 Phys.\ Rev.\  D {\bf 80} (2009) 015007
 [arXiv:0905.1148 [hep-ph]]. 








\end{thebibliography}
\end{document}